\pdfoutput=1

\documentclass[11pt]{article}
\usepackage{jheppub}

\usepackage{amsmath}
\usepackage{latexsym,amssymb}
\usepackage{graphicx,color,slashed}
\usepackage{bbm} 
\usepackage{mathtools}
\usepackage{ifpdf}

\newcommand{\cG}{\mathcal{G}}

\newcommand{\cH}{\mathcal{H}}

\newcommand{\cN}{\mathcal{N}}

\newcommand{\cV}{\mathcal{V}}

\newcommand{\be}{\begin{equation}}
\newcommand{\ee}{\end{equation}}
\newcommand{\ba}{\begin{eqnarray}}
\newcommand{\ea}{\end{eqnarray}}

\renewcommand{\d}{\textrm{d}}

\renewcommand{\a}{\alpha}

\newcommand{\N}{\mathcal{N}}

\def\E{{$E_{7(7)}$}}

\newcommand{\rf}[1]{(\ref{#1})}
\newcommand{\bea}{\begin{eqnarray}}
\newcommand{\eea}{\end{eqnarray}}

\def\bfzero{\relax{\rm I\kern-.18em 0}}
\def\bfone{\relax{\rm 1\kern-.35em 1}}
\def\twomat#1#2#3#4{\left(\begin{array}{cc}
\end{array}
\right)}

\def\d{\delta}

\title{\rm{\bf     Soft Theorems, Anomalies and Precursors}}
\author
{ Renata Kallosh}
 \affiliation{Stanford Institute for Theoretical Physics and Department of Physics,\\
  Stanford University, Stanford, CA 94305, USA}

\abstract{   We find that in 6d maximal supergravity  there are 3-loop order amplitudes with non-vanishing soft-scalar limits and anomalous $E_{5(5)}$ duality, induced by a 3-loop UV divergence. If the relevant non-vanishing soft-scalar limits would be  detected in 1-loop amplitudes, it would be  a  1-loop precursor of a 3-loop UV divergence in 6d. 
}

\begin{document}

\maketitle

 %\tableofcontents{}

%\newpage

\parskip 5pt

%%%%%%%%%%%%%%%%%%%%%%%%%%%%%%%%%%%%%%%%%%%%%%%%%%%%%%%%%%%%%%%%%%%%%%
%%%%%%%%%%%%%%%%%%%%%%%%%%%%%%%%%%%%%%%%%%%%%%%%%%%%%%%%%%%%%%%%%%%%%%
%%%%%%%%%%%%%%%%%%%%%%%%%%%%%%%%%%%%%%%%%%%%%%%%%%%%%%%%%%%%%%%%%%%%%%
%%%%%%%%%%%%%%%%%%%%%%%%%%%%%%%%%%%%%%%%%%%%%%%%%%%%%%%%%%%%%%%%%%%%%%
%%%%%%%%%%%%%%%%%%%%%%%%%%%%%%%%%%%%%%%%%%%%%%%%%%%%%%%%%%%%%%%%%%%%%%

\section{Introduction}

$\cN\geq 5$-extended supergravities in 4d were argued to be UV finite if $\cG$ duality symmetries, local $\cH$ symmetry and nonlinear  local supersymmetry have no anomalies \cite{Kallosh:2018wzz,Gunaydin:2018kdz,Kallosh:2023asd,Kallosh:2023thj}. Therefore it is important to identify and study all cases where some of these symmetries could be anomalous.

Known UV divergences in $\cN\geq 4$ supergravities include the case in 4d of a 4-loop UV divergence in half-maximal supergravity  \cite{Bern:2013uka}. In 6d in maximal supergravity there is a  3-loop  UV divergence   \cite{Bern:2008pv}.    
It is known that $\cN=5$ is UV finite at $L=\cN-1=4$ due to a cancellation of the 82 4-point diagrams \cite{Bern:2014sna}. The status of UV divergences in higher $\cN$ supergravities and higher loop orders is not known and the relevant computations are not likely to take place soon.

Is it possible to find some precursors for higher loop UV divergences by performing some  computations which in known cases of 4d $\cN=4$ and 6d $\cN=8$ can be tested to show a predictive power? 
For example, in 4d $\cN=4$ supergravity the $U(1)$ $\cH$-anomaly was found in  \cite{Marcus:1985yy} and anomalous $U(1)$ 1-loop amplitudes were also found in \cite{Carrasco:2013ypa}. Also, certain 1-loop $\cN = 4$ supergravity amplitudes that have non-vanishing soft-scalar limits were found in  \cite{Carrasco:2013ypa}, which means that $SU(1,1)$  duality $\cG$ symmetry was also broken at a 1-loop order.

In the case of 6d $\cN=8$ supergravity with the 3-loop UV divergence, such a detailed analysis of amplitudes and their anomalies was not performed. Moreover,   maximal 6d supergravity, according to  \cite{Marcus:1985yy} is anomaly-free: namely the $\cH$ symmetry has no 1-loop anomalies \cite{Marcus:1985yy} in
6d $\cN=8$ supergravity. But there is a 3-loop UV divergence, and there is no known 1-loop precursor, which is puzzling. 
In particular, the soft-scalar limits in multi-point amplitudes of this theory were not studied in detail. Maybe the relevant $\cG$ duality symmetry has anomalies?

In general, $\cN$-extended supergravities have physical scalars in $ {\cG\over \cH}$ coset space. Here 
$\cG$ is a noncompact Lie group and $\cH$ is its maximally compact subgroup. 
The issue of UV divergences in $\cN$-extended supergravities was studied from the perspective of $\cG$ duality symmetry in \cite{Broedel:2009nsh,Beisert:2010jx} and in \cite{Carrasco:2013ypa,Huang:2015sla,Freedman:2018mrv}. In the linear approximation, $\cG$ duality implies that the super-amplitudes should vanish in the soft scalar limit.

In particular, it was established in \cite{Beisert:2010jx,Freedman:2018mrv} that if there would be 4-point UV divergences at $L\leq \cN-2$   loop order in 4d supergravities, the soft scalar limit would not vanish at the 6-point  amplitudes at the same loop order\footnote{Analogous argument is not valid in  $L= \cN-1$   loop order. For example, $\cN=5, L=4$ UV finiteness is not explained by the soft scalar limit analysis \cite{Freedman:2018mrv}.}. For example, in the case of  $\cN=8$ with $\cG= E_{7(7)}$, the prediction of unbroken \E\ symmetry is that there should be no UV divergences at $L\leq 6$.  This explained the UV finiteness at 3-loop, and it remains a prediction of  UV finiteness for $L\leq 6$ if \E\, symmetry has no anomalies. 
In $N=5$ with $\cG= U(1,5)$ the UV finiteness  at $L=3$ is explained  \cite{Freedman:2018mrv}.  For $\cN=6$  with $\cG= SO^*(12)$ the UV finiteness is predicted for $L=3,4$  \cite{Freedman:2018mrv}.

In \cite{Carrasco:2013ypa,Huang:2015sla} the soft behavior of the multi-point amplitudes was studied in half-maximal 4d supergravity. It was suggested there following  \cite{Bern:1998sv} that the soft scalar function relating $n$-point amplitude to the $n-1$ point amplitude does not receive loop corrections. It was also stressed in  \cite{Huang:2015sla} there that 
  the soft scalar limit is not only non-vanishing for $U(1)$ non-preserving amplitudes, but for $U(1)$ preserving ones as well.

The maximal (2,2) supergravity in 6d \cite{Tanii:1984zk,Bergshoeff:2007ef} is characterized by a  coset space
$\cG\over \cH$, where $\cG=E_{5(5)}\sim SO(5,5)$ and $\cH= SO(5)\times SO(5)\sim {\rm USp(4)}\times {\rm USp(4)}$. 

The purpose of this work is to study the soft scalar limit of 6-point amplitude, which is required for  a nonlinear completion of the 4-point 3-loop UV divergence. We will find out, as it was the case for $L\leq \cN-2$  in 4d supergravity in \cite{Beisert:2010jx,Freedman:2018mrv}, that the soft scalar limit of 6-point amplitude does not vanish since the corresponding local superinvariant preserving R-symmetry is not available. This means that the 3-loop UV divergence breaks $\cG=E_{5(5)}$ duality symmetry. 

This observation is in agreement with a more general statement in \cite{Kallosh:2023css} that the 3-loop UV divergence is one loop order below the critical one, $L_{cr}=4$,  where the geometric superinvariants preserving $\cG$ symmetry and local $\cH$-symmetry exist.
The non-vanishing soft scalar limit of 6-point amplitude is a particular example demonstrating the fact that the 3-loop UV divergence in maximal (2,2) 6d supergravity necessarily breaks $\cG=E_{5(5)}$ symmetry, as well as a local $\cH=USp(4)\times USp(4)$ symmetry.

Our result about the 3-loop anomaly in $\cG=E_{5(5)}$ duality symmetry might have some relation to studies of differential equations satisfied by the string theory threshold functions, which have some anomalous terms induced by 3-loop UV divergences in 6d supergravity, \cite{Green:2010wi,Pioline:2015yea}.  

%%%%%%%

Given that the 3-loop 6-point amplitude associated with the 3-loop UV divergence has a non-vanishing soft scalar limit, we will discuss the possibility that some  1-loop multi-point amplitudes might also have a non-vanishing soft scalar limit. 
 The fact that at the 3-loop  $E_{5(5)}$ symmetry is broken by UV divergence suggests that the analysis of the 1-loop soft scalar limits might be very interesting.

 If the 1-loop $n$-point amplitudes have a vanishing soft scalar limit, it will be a case of a pure supergravity where the properties of the soft scalar functions are not universal: they depend on loop order, contrary to standard expectations.
 If the relevant 1-loop $n$-point amplitudes have a non-vanishing soft scalar limit, it will make them precursors of the 3-loop UV divergence.

\section{Maximal (2,2) 6d supergravity: amplitudes and superinvariants}\label{Sec:2}
\subsection{3-loop 4-point UV divergence}
The   (2,2) nonlinear local supersymmetry, local $\cH= SO(5)\times SO(5)\sim  {\rm USp(4)}\times {\rm USp(4)}$ symmetry and global $\cG=E_{5(5)}\sim SO(5,5)$  symmetry are properties of the  the classical action \cite{Tanii:1984zk,Bergshoeff:2007ef}. In the unitary gauge where the local $\cH$-symmetry is gauge-fixed, there are  25 physical scalars, coordinates of the  $\cG\over \cH$ coset space.

Using superapmlitudes, the structure of the maximal supergravity 4-point tree amplitude was given in  \cite{Cachazo:2018hqa}. The corresponding  on shell superfield   depends on 8 Grassmann coordinates, $\eta^{I, a}$, $\tilde \eta ^{\hat I, \hat a}$
\be
\Phi(\eta)
=  \phi + \ldots +  \eta_a^I \, \eta_{b,I} \, \tilde{\eta}^{\hat I}_{\hat{a}} \,\tilde\eta_{\hat b, \hat I}\, G^{ab;\hat{a}\hat b} + \ldots
+ (\eta)^4 (\tilde{\eta})^4 \bar\phi \,. 
\label{Ca}\ee
Here $I=1,2$ and ${\hat I}={\hat 1}, {\hat 2}$ label components of an $\text{SU}(2) \times \text{SU}(2)$ subgroup of the R symmetry group. Only this subgroup of the $\text{USp}(4) \times \text{USp}(4)$
R symmetry is manifest in this formulation. The on-shell field $G^{ab;{\hat a}{\hat b}}$ in the middle of the on-shell superfield is d=6 graviton, $a, \hat a=\pm$ are little-group indices. Two of the 25 scalar fields are in the 1st and last component of this superfield; the other 23 are in some middle components.

The four-point superamplitude is given in \cite{Cachazo:2018hqa} in the form
\be \label{eq:sugra-4pts}
{M}_4^{\N=(2,2)\text{ tree}} \;=\; {1\over \kappa^2} \d^6 \left( \sum_{i=1}^4 p_i^{AB} \right) { \delta^{8} \left( \sum_{i=1}^4 q^{A, I}_i \right)  \delta^{8} \left( \sum_{i=1}^4 {\tilde q}^{ {\hat I} }_{i,{\hat A}}\right)  \over s_{12}\, s_{23}\, s_{13} } \, ,
\ee
which has manifest permutation symmetry. Here the supercharges are defined as $q^{A, I}_i = \lambda^A_{i, a} \eta^{I, a}_i$, and $  {\tilde q}^{ {\hat I} }_{i,{\hat A}}= \tilde{\lambda}_{i, \hat{A},\hat{a} } \tilde{\eta}^{ {\hat I}, \hat {a}}_i$. These are half of the supercharges, and the other half involve $\eta$ derivatives. Conservation of these additional supercharges automatically follows from the first set together with the R symmetry.

The on-shell superfield \rf{Ca} is not R-symmetry invariant, therefore the amplitudes obtained in this formalism are not manifestly symmetric under  $\text{USp}(4) \times \text{USp}(4)$. However, one can define\footnote{C. Wen, private communication} a different on-shell superfield, which is neutral under R-symmetry, but transforms non-trivially under the little group. This is achieved by a Grassmann Fourier transform. In this way, one can prove that the 4-point amplitude has full $\text{USp}(4) \times \text{USp}(4)$ R-symmetry,  not just its manifest subgroup $\text{SU}(2) \times \text{SU}(2)$.

We can present a local linearized superinvariant defining  3-loop 4-point UV divergence found in \cite{Bern:2008pv} as follows
\be \label{3L4pts}
{M}_4^{\N=(2,2)\text{ L=3}} \;=\; {1\over \epsilon} {5\zeta_3\over (4\pi)^9} \Big ({\kappa\over 2}\Big)^4  \d^6 \left( \sum_{i=1}^4 p_i^{AB} \right) { \delta^{8} \left( \sum_{i=1}^4 q^{A, I}_i \right)  \delta^{8} \left( \sum_{i=1}^4 {\tilde q}^{ {\hat I} }_{i,{\hat A}}\right)   s_{12}\, s_{23}\, s_{34}     } \, ,
\ee 
It is proportional to ${M}_4^{\N=(2,2)\text{ tree}}$ and the difference in  $\kappa^6$ between 3 loops and tree is taken care by a factor $(s_{12}\, s_{23}\, s_{13} )^2$, since dimension of $\kappa^6$ equals -12 here. The numerical factors in eq. \rf{3L4pts} are taken from \cite{Bern:2008pv}  where this 3-loop UV divergence was computed.

 It is also possible, in direct correspondence with eq. \rf{3L4pts}  to present the 4-point UV divergent   amplitude using 1/2 BPS superinvariant. 
   \be
{M}_4^{\N=(2,2)\text{ L=3}} \quad \rightarrow \quad \kappa^4 \int d^6 x  [D_3 D_4 D^{\hat 3} D^{\hat 4} ]^4 \partial_\mu \partial_\lambda W \partial^\mu \partial_\nu W \partial^\nu \partial^\lambda W \,  W \ ,
\label{L3}  \ee
or in momentum space
   \be
\kappa^4  \delta( \sum_{m=1}^{4} p_m) \int  [D_3 D_4 D^{\hat 3} D^{\hat 4} ]^4  W(p_1, \theta)  W(p_2, \theta)  W(p_3, \theta)  W (p_4, \theta) s_{12}\, s_{23}\, s_{13}  \ ,
\label{L3m}  \ee
where the 1/2 BPS superfield is
\be
W \equiv W_{12}^{\hat 1 \hat 2} \ .
\label{W}\ee 
Here, in general,  the scalar superfield is $W_{ij}^{\hat i \hat j}$ with $i,j,\hat i, \hat j = 1, 2, 3, 4$, but in the linear approximation $ W_{12}^{\hat 1 \hat 2}$ is 1/2 BPS superfield depending on half of fermionic directions in superspace
\be
D_{\a 1} W_{12}^{\hat 1 \hat 2} = D_{\a 2} W_{12}^{\hat 1 \hat 2} = D^{\hat \a \hat 1} W_{12}^{\hat 1 \hat 2} D^{\hat \a \hat 2} W_{12}^{\hat 1 \hat 2}=0   \ .
\label{12BPS}\ee 
The 3-loop UV divergence in \rf{3L4pts}, \rf{L3m} is a linearly supersymmetric version of a gravitational part of it. In  symbolic form, it is
\be
\kappa^4 \int d^6 x \, D^6 R^4 +\dots
\ee
As different from \cite{Bossard:2009sy}, where related linearized harmonic superinvariants 
were presented, 
we are not using  in \rf{L3}, \rf{L3m} harmonic superspace coordinates to break  R symmetry: the  choice of the 1/2 BPS superfield requires such a breaking anyway. Namely, $\text{USp}(4) \times \text{USp}(4)$ can be  manifestly broken down to $\text{SU}(2) \times \text{SU}(2)$  or with the help of harmonic coordinates.
 However, in both cases, the 4-point amplitude is invariant under $\text{USp}(4) \times \text{USp}(4)$ due to the combined properties of the measure of integration and Lagrangian.

One can see why the 4-point linearized superinvariant in \rf{L3m} has an unbroken global R-symmetry. The measure of integration and the Lagrangian can be given in the manifestly R-invariant form
 \be
\Big ([D_3 D_4 D^{\hat 3} D^{\hat 4} ] \,  ( W_{12}^{\hat 1 \hat 2}) \Big )^4 s_{12}\, s_{23}\, s_{13}\rightarrow \Big ([D_i D_j D^{\hat i} D^{\hat j} ] \,  ( W_{kl}^{\hat k \hat l}) \epsilon ^{ijkl} \epsilon _{\hat i \hat j \hat k\hat l}\Big )^4s_{12}\, s_{23}\, s_{13}
\label{balance} 
\ee

\subsection{No local 6-point  superinvariant with non-vanishing SSL at 3-loop order}

 To build superinvariants representing multi-point amplitudes we cannot use the 1/2 BPS superfields \rf{W} anymore. 
If we would try to have more than four of the $W_{12}^{\hat 1 \hat 2}$ superfields, we would break R-symmetry of the action: the balance between the measure of integration and the Lagrangian valid for the 4-point linear superinvariants in \rf{balance} would be lost \footnote{An additional way to see that $n$-point  superinvariants  constructed from 1/2 BPS superfields break R-symmetry is that these do not have R-invariant pure gravity sector for $n>4$. Having pure gravity sector  is only a property of $n=4$ BPS superinvariants. This fact is known from the time the first BPS superinvariants were constructed in \cite{Kallosh:1980fi,Howe:1981xy}.}.  

There are more 4-point harmonic superinvariants in \cite{Bossard:2009sy}, depending on 1/4 and 1/8 BPS superfields, which require specific choices of scalar superfield directions $W_{kl}^{\hat k \hat l}$. For example, a quartic combination of $W_{1r}^{\hat k \hat l}$ where $r=3,4$ allows a 1/8 BPS superinvariant: it is an integral over 28 out of 32 fermionic directions.
We have checked that all of these BPS superinvariants do not have generalization to $n$-point superinvariants with $n>4$, they always break R-symmetry. 

Thus, to describe the local multi-point linearized superinvariants, we need to use the unconstrained superfields $W_{ij}^{\hat i \hat j}$ and integrate over the whole superspace. An analogous experience was encountered in the past in 4d where linearized BPS invariants are only available for 4-point amplitudes, see for example \cite{Freedman:2018mrv}.

We will now employ to maximal 6d supergravity the strategy used in \cite{Beisert:2010jx,Freedman:2018mrv}  to find local 6-point invariants which might cancel the 6-point completion of the UV divergences in 4d $\cN\geq 5$ supergravities.

The superfield $W_{kl}^{\hat k \hat l}$ has 25 scalars in its 1st component:  $ W_{kl}^{\hat k \hat l}$  can be given in the form \be
W_{kl}^{\hat k \hat l}=  \gamma^{a}_{ kl}V_a^{\hat a}  \gamma_{\hat a}^{ \hat k\hat l} \ ,
\ee
 where $a, \hat a =1,2,3,4,5$ and $V_a^{\hat a}$ has 25 independent components. Here $a, \hat a$ are $SO(5)\times SO(5)$ vector indices, whereas $i,j, \hat i, \hat j=1,2,3,4$ are spinorial ones
\footnote{More details can be found in Appendix in the context of d=6 maximal supergravity action \cite{Tanii:1984zk,Bergshoeff:2007ef}. There is a small difference in notation: in amplitude notation, we use here (with an account 
 of \cite{Cachazo:2018hqa,Bossard:2009sy})
the second $ \text{USp}(4)\sim SO(5)$ indices are $\hat \a, \hat \beta$ and $\hat a, \hat b$ whereas in \cite{Tanii:1984zk,Bergshoeff:2007ef} the second  $ \text{USp}(4)\sim SO(5)$ indices go with $\dot \a, \dot  \beta$ and $\dot a, \dot b$. Also spinorial $i,j,\hat i, \hat j = 1, 2, 3, 4$ are in \cite{Tanii:1984zk,Bergshoeff:2007ef}  $\a,\beta,\dot \a, \dot \beta = 1, 2, 3, 4$.}.

Under linearized $E_{5(5)}\sim SO(5,5)$ symmetry, the linearized scalar superfield is shifted by a constant, which is the coset type part of this symmetry
\be
\delta_{\cG} V_a{}^{\hat a}(x, \theta_{\pm}) = \Sigma_a{}^{\hat a}  \ .
\label{d6soft1} \ee
The 6-point candidate 3-loop linearized superinvariants are
\be
S^6= \kappa^4  \int d^6 x \,  d^{32} \theta \,  Tr \, V^6 \ ,
\label{L36p}  \ee
where $Tr \, V^6$ means any $SO(5)\times SO(5)$-invariant combination of 6 superfields $W_a^{\hat a}$.  Here the metric $\delta^{ab}$ and $\delta _{\hat a \hat b}$ can be used. Under the linearized coset part of $E_{5(5)}\sim SO(5,5)$ symmetry \rf{d6soft1} the action transforms as follows
\be
\delta_{\cG} S^6= 6 \, \kappa^4  \int d^6 x \,  d^{32} \theta \,  Tr ( \Sigma \, V^5 )\neq 0 \ .
\label{L36pdelta}  \ee
Therefore  $E_{5(5)}$ is broken by $S^6$.
Thus, if the 6-point local superinvariant is available at 3-loop order, it breaks the soft scalar limit. This is what is needed to cancel the non-linear completion of the 4-point 3-loop UV divergence.

The next step is the one we made in \cite{Freedman:2018mrv} to establish when the local superinvariants are available and when they are not available, i. e. we need a dimensional analysis in superspace. 
Using the same technique, we have found that in 4d  $\cN\geq 5$, the 6-point local amplitudes with a non-vanishing SSL are available only starting from $L=\cN-1$. At  $L\leq\cN-2$ these are not available\footnote{In \cite{Freedman:2018mrv} in addition to performing the 6-point superinvariant analysis, we have also computed the relevant 6-point amplitudes, which was a significantly more complicated process. The result of amplitude computations fully confirmed the analysis based on superinvariants. In both cases, linearized supersymmetry was preserved. Now in 6d maximal supergravity, we have studied here the superinvariants in eq. \rf{L36p}.}.
Therefore the 4-point UV divergence at $L\leq \cN-2$ requires a nonlinear completion with a non-vanishing soft scalar limit, and there is no extra local 6-point superinvariant to cancel it. The conclusion was that in 4d the UV divergence at $L\leq \cN-2$ leads to a 6-point completion breaking E7 type symmetry.

Here we have the following dimension counting in eq. \rf{L36p}
\be
-8 -6+16 =2\neq 0
\ee
There is a contribution from  $\kappa^4$, from six dimensions, from 32 supersymmetries, and we note that  the superfields $V^a_{\hat a}$ are dimensionless. The total is 2, whereas we need an action to have dimension 0.
Thus, at 3 loops the local 6-point superinvariant \rf{L36p} is unavailable. It means there is no way to cancel the soft scalar limit, which is non-vanishing due to 4-point 3-loop UV divergence $D^6 R^4+\dots $ in \rf{3L4pts}, \rf{L3m} when this linearized expression is supplemented by nonlinear completion.

This is precisely the argument in \cite{Freedman:2018mrv}, which explains why a potential  6-loop $\cN=8$ UV divergence $D^6 R^4$ would mean that \E\, is broken. There, in eq. (4.6)  the symbolic candidate for a 6-point local linearized superinvariant was
\be
S^6= \kappa^{2(\cN-2)}  \int d^4 x \,  d^{4\cN} \theta \,  Tr \, W^6  \ .
\ee 
The counting was for $L=\cN-1$
\be
-2(\cN-2) -4+2\cN=0
\ee 
But $S^6$ for any loop below $L=\cN-1$   has a positive dimension, and the relevant local superinvariant is unavailable.

There is an obvious similarity between the 6d supergravity and the 4d one. Also, in both cases, the absence of a local 6-point amplitude with a non-vanishing soft scalar limit extends to any $n>6$ point amplitude.

\subsection{ $L\geq L_{cr}= 4$} 

Thus, the 4-point 3-loop UV divergence discovered in \cite{Bern:2008pv} can be described in a form with linearized (2,2) supersymmetry, which does not have a nonlinear generalization at the 3-loop order. The ones that have nonlinear supersymmetry start at the 4-loop order and are given by a geometric whole superspace integral 
\be
\kappa^{2(L-1)} \int d^6x \, d^{32} \theta \, {\rm Det} \, E\, {\cal L} (T, R)\, ,\qquad L\geq 4 \ ,
\ee
where  $T, R$ are superspace torsions and curvatures, which are $\cG$-invariant. The dimension of 
 ${\cal L}$ is  $ 2+n$ , $n>0$ as it depends on a product of at least 4 superspace torsions and curvatures. The smallest is a  spinorial superfield of dimension 1/2, a superspace torsion. Dimension counting goes like this
\be
-4(L-1)-6+16+2+n=0 \ ,
\ee
which means
\be
L= 4+{n\over 4}\, , \qquad n>0 \ .
\ee
Geometric superinvariants by construction have an unbroken $\cG$-symmetry, and in the linear approximation, it means that soft scalar limits are vanishing.  The gravitational part is 
\be
D^{10+n} R^4 \qquad n>0 \ .
\ee
The total number of derivatives is $18+n$.

Clearly, at the 3-loop order with $D^6 R^4$ and a total number of derivatives 14, we have only linearized non-geometric superinvariants depending on scalar superfields of dimension zero; they break $\cG$ symmetry in the form of non-vanishing single scalar limit.

The  question is, was it necessary to perform a 3-loop computation of the 4-point UV divergence, or there are  some 1-loop precursors?

\subsection{Relation to IIB string theory non-renormalisation theorems}
  A related study consistent with our results concerning 6d maximal supergravity was performed
in the context of IIB superstring theory in  \cite{Wang:2015jna}, \cite{Green:2019rhz}. In these studies, various non-renormalization theorems for higher derivative terms arising from $\alpha' $ expansion were derived using superamplitudes.

Our main result is that one cannot construct R-symmetry invariant local terms at higher points with $\leq 14$ derivatives in 6d maximal supergravity. In terms of 10d superamplitudes using 10-dimensional helicity spinors, a related observation was made in \cite{Green:2019rhz}. The linearized BPS superfields in 10d depend on 8 $\eta$'s, and there are 2 complex scalars, $Z$ and $\bar Z$.

An example of the  6-point superamplitude studied in  \cite{Green:2019rhz} in detail has a term with 4 gravitons, one scalar $Z$, and one $\bar Z$. It has a form
\be
\delta^{16} (Q_n) (\bar Q_n)^{16} (\eta_i)^8 (\eta_j)^8 (\eta_k)^8 \ ,
\ee
where $Q_n^A= \sum_{i=1}^n \lambda_{i, a}^A \eta^a_i$ and $\bar Q_n^B= \sum_{i=1}^n \lambda_{i}^{B, a}  {\partial \over \partial \eta^a_i}$. Here $A = 1,...,16$ labels the components of a $SO(9,1)$ chiral spinor, and $a = 1,...,8$ labels the components of a $SO(8)$ spinor of the little group of massless states. Thus, there are 16 powers of momentum, whereas the relevant BPS  have at most 14 powers. The conclusion here is that 
 in order to describe a supersymmetric term, there must be intermediate poles (inverse momentum factors), so supersymmetric contact terms are not allowed. In fact,  more examples  which are consistent with $U(1)_R$ symmetry for IIB string theory were studied in \cite{Green:2019rhz}. They have also shown 
 that it is not possible to construct the relevant superamplitudes without poles.
 
The conclusion that supersymmetric contact terms of dimension $\leq 14$ are not allowed for non-maximal U(1)-violating processes in 10d supergravity can be transferred to 6d supergravity. In 10d we have ${\cG\over \cH}= {SL(2,\mathbb{R})\over U(1)}$ coset space with 2 scalars, in 6d there is $ {\cG\over \cH}= {E_{5(5)}\over USp(4)\times USp(4)}$  coset space with 25 scalars. The soft scalar limits are defined by $\cG$ symmetries in all cases.
The conclusion described above in the context of type IIB string theory is consistent\footnote{C. Wen, Yu-tin Huang, private communication.} with our findings here using superfields in 6d supergravity where the relevant duality symmetry is $E_{5(5)}$.

\section{Discussion}

$ \cG$ and $ \cH$ symmetries of nonlinear $\cN$-extended supergravity with physical scalars in ${\cG\over \cH}$ coset space can be tested in the linearized approximation using computations of quantum corrections to on shell super-amplitudes.
In the linear approximation to supergravity, there is a global $\cH$-symmetry. The footprint of duality  $\cG$-symmetry is a requirement of a vanishing soft scalar limit.

In 4d in $\cN\geq 4$ pure supergravities so far, the   UV divergences were found only at $L=4$ in $\cN=4$ case \cite{Bern:2013uka}. At the 1-loop order, this theory has $\cH$-symmetry $U(1)$ anomaly \cite{Marcus:1985yy,Carrasco:2013ypa,Huang:2015sla} as well as non-vanishing soft scalar limits in multipoint superamplitudes. These non-vanishing soft scalar limits are $\cG$ symmetry anomalies.
Thus, in the case of 4d  $\cN= 4$  supergravity, the 4-loop UV divergences have  1-loop precursors, both $\cH$ as well as $\cG$ symmetry anomalies.

In 6d maximal supergravity, there is a 3-loop UV divergence  \cite{Bern:2008pv}. It was explained in \cite{Kallosh:2023css} that the 3-loop UV divergence, being non-geometric, breaks a nonlinear $\cG$ symmetry as well as a local $\cH$-symmetry. 3-loop UV divergence is one loop order below the critical one, $L_{cr}=4$,  where the geometric superinvariants, preserving $\cG$ symmetry and local $\cH$-symmetry exist.

Here we studied the effect of the presence of a 3-loop UV divergence in maximal 6d supergravity on the linearized  $\cH$ and $\cG$ symmetries. These issues  can be explored 
 using linearized, non-geometric superinvariants, or superamplitudes, the way it was done for 4d supergravities in \cite{Freedman:2018mrv}.
 
 The main result here is that there are no 6-point local superamplitudes at L=3 with a non-vanishing scalar soft limit, which would cancel the nonlinear 6-point completion of the 4-point UV divergence. Thus, at the linear level, 3-loop UV divergence breaks $\cG=E_{5(5)}$  since the soft scalar limit of the 6-point superamplitude is not vanishing.
 
 Here we have used only the method of linearized superinvariants\footnote{The analysis of the nonlinear on shell superspace was performed in \cite{Bossard:2014lra}. The harmonic space superinvariants are required for $L\leq 3$ loop order. It was, however, found in \cite{Bossard:2014lra} that  the harmonic measure of integration does not exist at the nonlinear level, and the underlying G-analiticity condition  admits obstructions. This is consistent with our result on $E_{5(5)}$ anomaly induced by the 3-loop UV divergence in maximal 6d supergravity.}.
 In an analogous study in \cite{Freedman:2018mrv}, we constructed linearized superinvariants as well as 6-point superamplitudes satisfying supersymmetric Ward identities. The 6-point superamplitude computation confirmed the result following from superinvariants. In 6d case a more recent analysis in  \cite{Kallosh:2024lsl} has shown  that the  superamplitudes, satisfying supersymmetric  Ward identities,  must give the same result as the one from the analysis of linearized superinvariants performed in this paper.

It would be interesting to study  1-loop  $n$-point superamplitudes.  The global $\cH=USp(4)\times USp(4)$-symmetry  does not have 1-loop anomalies \cite{Marcus:1985yy}, but the status of the soft scalar limit is not known, as far as we understand.

There are two possible outcomes of such a potential study.

1. 1-loop $n$-point super-ampiltudes in maximal 6d supergravity have a vanishing  soft scalar  limit. This  would mean that the soft scalar functions are not universal, they vanish at 1-loop order but do not vanish at the 3-loop order. One expects that these functions are loop-order independent, see the discussion of this in \cite{Bern:1998sv,Carrasco:2013ypa,Huang:2015sla}.

2. 1-loop $n$-point superampiltudes in maximal 6d supergravity have a non-vanishing  soft scalar  limit.  It would be a precursor of the 3-loop 4-point UV divergence detectable via 1-loop order computations of multi-point superamplitudes.

If option 2  is realized, one would be able to study the multi-point 4d $\cN\geq 5$ supergravity superamplitudes and their soft scalar limits in the hope that these are precursors of higher loop UV divergences. So far, only the absence of $U(1)$ anomalies in 4d $\cN\geq 5$ supergravity superamplitudes was observed in \cite{Freedman:2017zgq}. However, not much is known about 1-loop multi-point soft scalar limits. But if 6d maximal supergravity has UV divergence at 3-loop and a $\cG$-anomaly precursor at 1-loop, it would suggest that one can look for 1-loop precursors of   UV divergences in supergravities where higher loop computations are extremely difficult.

\

\noindent{\bf {Acknowledgments:}} I am grateful to J. J. Carrasco, H. Elvang, D. Freedman, Y.-t. Huang, E. Ivanov, A. Linde, H.~Nicolai, R. Roiban, C. Wen and  Y. Yamada for stimulating discussions of the UV divergences in supergravities. 
 This work is supported by SITP and by the US National Science Foundation grant PHY-2310429.

\appendix 

\section{ Duality  $\cG$ symmetry and soft scalar limit approximation}
\subsection{Local $\cH=SU(8)$ symmetry in maximal d=4 supergravity}

The analysis of the soft  single and double scalar limits are based on the linear realization of both $\cG$ and $\cH$ symmetries where the Lie algebra $\mathbb{ G}$ is split
\be\label{split}
\mathbb{ G}= \mathbb{ H}  \oplus \mathbb{ K}\ee
Here $\mathbb{ H}$ is the Lie algebra of $\cH$. $\mathbb{ K}$
 contains the remaining generators referred to as ``coset generators''. The algebra is 
 \be\label{HK}
[ \mathbb{ H} ,  \mathbb{ H} ]\subset  \mathbb{ H} \, , \qquad [ \mathbb{ H} ,  \mathbb{ K} ]\subset  \mathbb{ K} \, , \qquad[ \mathbb{ K} ,  \mathbb{ K} ]\subset  \mathbb{ H} \ .
\ee
 The supergravity action has a local $\cH$ symmetry, when physical and unphysical scalars are present, and there is a global $\cG$-symmetry. These symmetries are independent and linearly realized; see, for example, $\cN=8, d=4$ action in \cite{Cremmer:1979up, deWit:1982bul}. 

At the nonlinear level, the scalar vielbein $\cV$ transforms under global $\cG$  and under local $\cH$ symmetries. When the local $\cH$ symmetry is gauge-fixed, the remaining scalars form coordinates of the coset space. In order to preserve this gauge, $\cG$  transformations must be accompanied by particular field-dependent $\cH$ transformations. The details are well known in d=4 $\cN=8$ case \cite{Cremmer:1979up, deWit:1982bul,Kallosh:2008ic,Bossard:2010dq}.

To understand these symmetries in the linearized approximation we consider  d=4 $\cN=8$ case with $ {\cG\over \cH}= {E_{7(7)}\over SU(8)}$, where the soft limits in amplitudes were investigated in \cite{Bianchi:2008pu,Arkani-Hamed:2008owk,Kallosh:2008rr}
whereas the compensating, preserving the unitary gauge, field-dependent $SU(8)$ transformation was studied in
detail in \cite{Kallosh:2008ic,Bossard:2010dq}.

 At the whole nonlinear level, scalars before gauge-fixing are in the $E_{7(7)}$ group element, there are 133 of them, defined by the 56-bein connecting the $E_{7(7)}$ anti-symmetric pair of indices $IJ$ with the anti-symmetric pair of $SU(8)$ indices, $ij$.
 \begin{eqnarray}\label{gaugex}
\cV=\left(
                                        \begin{array}{cc}
                                          u_{ij}{}^{IJ} & v_{ijKL} \\
                                          v^{klIJ} & u^{kl}{}_{KL} \\
                                        \end{array}
                                      \right)\ .
\end{eqnarray}
  Under the combined action of local $SU(8)$ and global \E\,  the 56-bein transforms as
\be\label{LR}
{\cal V} (x)\quad \rightarrow \quad {\cal V}' (x)= U(x) {\cal V} (x) E^{-1}\, ,  \qquad \qquad 
U(x)\in SU(8) \;, \;\; E\in E_{7(7)} 
\ee
where $E\in E_{7(7)}$ and it is in the fundamental 56-dimensional representation where
\begin{eqnarray}\label{E77}
  E= \exp G_{E_{7(7)}} 
                                       \ , \qquad G_{E_{7(7)}}=\left(
                                        \begin{array}{cc}
                                         \Lambda_{IJ}{}^{KL} & \Sigma _{IJPQ} \\
                                          \Sigma^{MNKL} & \Lambda ^{MN}{}_{PQ} \\
                                        \end{array}
                                      \right)\ .
\end{eqnarray}
The $E_{7(7)}$ Lie algebra requires that
$
\Lambda_{IJ}{}^{KL}= \delta_{[I}{}^{[K} \Lambda_{J]}{}^{L]}
$.
Here $\Lambda_{I}{}^{J}$ are the generators of the $SU(8)$ maximal subgroup of $E_{7(7)}$ and $
\Lambda_I{}^J= - \Lambda ^J{}_I, \, \Lambda ^I{}_I=0
$.
With $\mathbb{ G}$ split as in eq. \rf{split} the algebra is in eq. \rf{HK}. The exact value of structure constants for \E\, is given in eq. (B.2) in \cite{Cremmer:1979up}.

The local $SU(8)$ transformation $U(x)$ acts on the 56-bein from the left in eq. \rf{LR} and is {\it completely independent of the $E_{7(7)}$} transformation
\begin{eqnarray}\label{gauge}
U(x) = \exp G_{SU(8)} \ ,   \qquad
G_{SU(8)}(x) =\left(
                                        \begin{array}{cc}
                                         \delta_{[i}{}^{[k} \Lambda_{j]}{}^{l]}(x) & 0 \\
                                          0& \delta^{[m}{}_{[p} \Lambda^{n]}{}_{q]}(x)\\
                                        \end{array}
                                      \right)\ .
\end{eqnarray}

\subsection{Unitary gauge with local $\cH=SU(8)$ symmetry gauge-fixed}
In the unitary gauge  \cite{Cremmer:1979up, deWit:1982bul,Kallosh:2008ic,Bossard:2010dq} there is no distinction between the \E\, indices $IJ$ and $SU(8)$ indices $ij$ and there are 70 scalars $\phi_{ijkl}=\pm  {1\over 4!}  \epsilon_{ijklmnpq} \bar \phi^{mnpq}
$
 \be
{\cal V}(x) ={\cal V}^{\dagger} (x) =
\exp\left( \begin{array}{cc}  0  &  \phi_{ijkl}(x) \vspace{0mm}  \\
 \bar \phi^{ijkl}(x)  & 0 \end{array} \right)\, .
\label{gauge1}\ee 
The 133-70=63 local parameters of $SU(8)$ were used to eliminate the unphysical 63 scalars. In the unitary gauge the remaining symmetries acting on scalars are conveniently given in \cite{Kallosh:2008ic} in terms of inhomogeneous coordinates of the ${E_{7(7)}\over SU(8)}$ coset space  
\be\label{yphi}
y_{ij, kl}(x)\equiv  \phi_{ijmn}(x)\left({\tanh(\sqrt {  \bar\phi \phi(x)}\over \sqrt {\bar  \phi \phi(x)}} \right)^{mn}_{\; \; \; \;\; {kl}} \ .
\ee
In terms of the nonlinear scalar fields $y, \bar y$,  the matrix $\cV$ is given by the following expression
\begin{eqnarray}\label{Vy}
\cV(y, \bar y)=\left(
               \begin{array}{cc}
                 P^{-1/2} & -P^{-1/2}y \\
                 -\bar{P}^{-1/2}\bar{y} & \bar{P}^{-1/2} \\
               \end{array}
             \right)\, , \qquad P(y,\bar{y})_{ij}{}^{kl}\equiv(\delta_{ij}{}^{kl} -y_{ijrs}\bar{y}^{rskl}) \ .
\end{eqnarray}
In the unitary gauge \E\, symmetry is non-linearly realized on scalars. A simple form of the exact nonlinear $E_{7(7)}$  transformation  on scalars  \rf{yphi} was presented in \cite{Kallosh:2008ic}
\begin{eqnarray}\label{deltay}
\delta y=y'-y= \Sigma+ y\bar{\Lambda}-\Lambda y-y\bar{\Sigma}y , 
\end{eqnarray} 
 In the linear approximation
\be\label{yphiLin}
y_{ij, kl}(x)=  \phi_{ijkl}(x) +\dots
\ee
Under $SU(8)$ subgroup in the linear approximation
\be
\delta  \phi (x) = \phi\bar{\Lambda}-\Lambda \phi
\ee
and under coset symmetries in the linear approximation
\be
\delta  \phi_{ijkl} (x)= \Sigma_{ijkl} +\dots
\ee
Thus, the non-compact transformations are nonlinear, with scalars transforming as Goldstone bosons to the lowest order. It is this coset symmetry of supergravity that can be tested in amplitude computations where the soft scalar limits can be computed.

The linearized version of the coset part of $\cG$ symmetry requires such soft scalar limits for a single scalar to vanish.

\subsection{6d maximal (2,2) supergravity with ${\cG\over \cH}={E_{5(5)}\over USp(4)\times USp(4)} \sim {SO(5, 5)\over SO(5) \times SO(5)}$} 

The action of maximal  supergravity in 6d \cite{Tanii:1984zk,Bergshoeff:2007ef} with the   coset space
is given in a form where there is a local $\cH= SO(5)\times SO(5)\sim {\rm USp(4)}\times {\rm USp(4)}$ symmetry.

There are 45 scalars defining the vielbein  
\be
{\cal V}_M{}^{\underline A}= \left(
                                        \begin{array}{cc}
                                          {\cal V}_{m}{}^{a} & {\cal V}_{m}{} ^{\dot a}\\
                                            {\cal V}^{ma} &  {\cal V}^{m \dot a} \\
                                        \end{array}
                                      \right)\  = \left(
                                        \begin{array}{cc}
                                          A & B\\
                                            C &  D \\
                                        \end{array}
                                      \right)   \ ,
\ee
where
\be
A^T C+ C^T A=1\, , \quad B^T D+ D^T B=-1 \, , \quad A^T D + C^T B=0 \ .
\ee
The index $\underline A= (a, \dot a)$  involves the vector indices of $SO(5) \times SO(5)$ and $M$ is in $SO(5,5)$.

In the unitary gauge, there are  25 scalar fields parametrizing the coset ${SO(5, 5)\over SO(5) \times SO(5)}$. In the linearized approximation these 25 scalars are in $V_a{}^{\dot a}$, where $a, \dot a$ are vector indices of $SO(5) \times SO(5)$, $a, \dot a=1,2,3,4,5$. The linearized superfield with the first component physical scalar $V_a{}^{\dot a}(x)$ is
\be
V_a{}^{\dot a}(x, \theta_{+ \a}, \theta_{-\dot \a} )
\ee
It depends on all 32 fermionic coordinates of the on-shell superspace. Here $ \theta_{+ \a}, \theta_{-\dot \a}$ are 8 symplectic Majorana-Weyl spinors and $\alpha, \dot \alpha =1,2,3,4$.

The coset generator symmetries related to soft scalar limits are, in the linear approximation
\be
\delta V_a{}^{\dot a}(x, \theta_{\pm}) = \Sigma_a{}^{\dot a} +\dots
\label{d6soft} \ee
To find the linearized 1/2, 1/4, 1/8  BPS superfields  depending on a fraction of $\theta$'s, one has to switch to superfields with spinorial indices in $USp(4)\times USp(4)$
\be
W_{\alpha \beta}^{\dot \alpha \dot \beta} \equiv  \gamma^a_{\alpha \beta} \gamma_{\dot a}^{\dot \a \dot \beta} V_a{}^{\dot a}
\ee
The BPS superfields we use in Sec. \ref{Sec:2} in eq. \rf{W} are $W_{12}^{\hat 1 \hat 2} $. Using exact local supersymmetry of the Lagrangian in \cite{Tanii:1984zk,Bergshoeff:2007ef} in linearized approximation, one finds the BPS constraints. For example, in eq. \rf{12BPS} we show that $W_{12}^{\hat 1 \hat 2} $ depends only on half of fermionic directions in superspace.

But since BPS superfields always break $USp(4)\times USp(4)$, one finds that restoring this symmetry is only possible for the 4-point linearized superinvariants due to special properties of the measure of integration.

In linearized multi-point superinvariants, it is necessary to get $USp(4)\times USp(4)$  invariant Lagrangians as well as a measure of integration. This makes choices we made in \rf{L36p}  the only possibility to study R-invariant amplitudes.

\bibliographystyle{JHEP}
\bibliography{refs}

\end{document}